**Charge transfer, band-like transport, and magnetic ions at $F_{16}$CoPc/rubrene interfaces**

*Yulia Krupskaya\*, Florian Rückerl, Martin Knupfer, and Alberto F. Morpurgo\**

Dr. Y. Krupskaya, Prof. A. F. Morpurgo
Department of Quantum Matter Physics (DQMP) and Group of Applied Physics (GAP),
University of Geneva, 24 quai Ernest-Ansermet, CH-1211 Geneva, Switzerland
E-mail: y.krupskaya@ifw-dresden.de, alberto.morpurgo@unige.ch
F. Rückerl, Prof. M. Knupfer
IFW Dresden, Helmholtzstr. 20, D-01171 Dresden, Germany

The phenomenon of charge transfer in molecular materials is of a particular interest both in fundamental and applied research in the field of organic electronics.[1, 2] Especially when considering interfaces formed by two different organic semiconductors, charge transfer from one semiconductor to the other can have dramatic effects and lead to interfacial electronic properties that differ drastically from the properties of the individual constituent materials.[1-18] Indeed, the interfaces between two large-gap, initially insulating organic semiconductors can exhibit significantly enhanced electrical conductivity[3-13] and in some cases even metallic behavior, as it has been observed in interfaces formed by single crystals of TTF (tetrathiofulvalene) and TCNQ (tetracyanoquinodimethane).[5] Another very interesting example is provided by recent experiments on interfaces formed by films of two different metal-phthalocyanine molecules (MnPc and $F_{16}$CoPc). Spectroscopic studies on this system have shown that charge transfer involves electrons that occupy orbitals centered on the magnetic metal ions, resulting in the formation of a new magnetic hybrid state at the interface.[16] Although only a small number of experiments have been performed until now, these examples suggest that organic semiconductors offer an unprecedented flexibility to control the electronic state of interfacial electronic systems.

Here we take a first step in realizing charge transfer interfaces that combine both a large electrical conductivity and the presence of magnetic ions. The ultimate goal of an electronic system combining these properties is to induce and control interfacial magnetism, with the charge carriers responsible for the enhanced interfacial conductivity also mediating the interaction between the magnetic ions, e.g., through double-exchange processes.[19] At the current stage, with no previous experiments reported, our work aims at identifying promising combinations of different molecular materials enabling the experimental realization of magnetic and conducting organic interfaces. This is essential because so far the number of conducting organic interfaces that have been investigated experimentally is limited,[3-13] in no cases involving magnetic ions, and among all systems studied only one was found to exhibit metallic interfacial electrical conductivity.[5]

We investigate heterostructures formed by a rubrene (tetraphenyltetracene) single crystal and an $F_{16}$CoPc (fluorinated Co-phthalocyanine) film by means of charge transport and photoelectron spectroscopy. We find that the $F_{16}$CoPc/rubrene interface has significantly enhanced electrical conductance. With the exception of TTF/TCNQ,[5] the room temperature resistivity of this interface is the lowest reported so far, and temperature dependent measurements show a decrease of the resistivity upon cooling (down to ~130 K) indicating the occurrence of band-like transport.[20-30] By means of Hall effect measurements we establish that charge transport is dominated by holes in the rubrene crystal and we find that the value of the hole mobility virtually coincides with that measured in a class of recently investigated organic interfaces based on $F_x$-TCNQ (fluorinated tetracyanoquinodimethane) and rubrene single crystals.[13] As compared to all interfaces studied in this family, on the contrary, the hole density is higher in $F_{16}$CoPc/rubrene, which explains the low value of the $F_{16}$CoPc/rubrene resistivity. The results of the transport measurements are corroborated by Kelvin probe microscopy experiments that enable the alignment of the chemical potential across the $F_{16}$CoPc/rubrene interface to be determined, and confirm that a large transfer between the two materials should be expected. Finally, we perform photoelectron spectroscopy (PES) measurements on thin film $F_{16}$CoPc/rubrene heterostructures to show that the charge transfer at the interface involves electronic orbitals centered on the magnetic Cobalt ion of the $F_{16}$CoPc molecules. We therefore conclude that the investigated system does allow the high electrical conductivity with the presence of magnetic ions to be combined at the interface.

$F_{16}$CoPc/rubrene interface devices were formed on a polydimethylsiloxane (PDMS) substrate. First a rubrene single crystal (grown by physical vapor transport) was laminated on the substrate and then a 70 nm $F_{16}$CoPc film was evaporated (under high vacuum conditions) on top of the rubrene crystal. The transport properties of rubrene single crystals (identical to those used to assemble the interfaces) were investigated via field-effect transistor (FET) measurements[31] as a characterization step, and showed a room temperature mobility between 12 and 20 cm$^2$V$^{-1}$s$^{-1}$ in devices in which the crystal was suspended on top of a gate electrode, in agreement with previous studies.[32-34] In order to maintain its quality, the rubrene crystal was kept at room temperature throughout the deposition of the evaporated film. As a result, the morphology of the $F_{16}$CoPc film was expected to be far from ideal, as indeed indicated by atomic force microscopy (AFM) measurements showing $F_{16}$CoPc films with rather rough surfaces, consisting of small grains with irregular orientation. Electrical contacts to the interface were realized manually using conducting carbon paste, following a strategy adopted earlier to perform transport measurements on different organic single crystal interfaces.[5] An optical microscope image of one of the devices investigated is shown in **Figure** 1a.

Figure 1b shows the I-V curve of a $F_{16}$CoPc/rubrene interface measured in vacuum using a multi-terminal device configuration (by means of an Agilent Technology E5270B parameter analyzer) and exhibiting linear characteristics. The measured conductance is many orders of magnitude larger than the conductance of the individual materials forming the interface. Specifically, the room temperature resistivity for all measured devices was found to be in the range of 260-350 kΩ/square. Hall effect measurements performed on the same devices (**Figure** 2a) show that charge transport in the $F_{16}$CoPc/rubrene interface is dominated by holes in rubrene crystals, as it may have been expected, since the charge carrier mobility in organic films is generally significantly lower than in crystals. Indeed, in our $F_{16}$CoPc/rubrene interfaces the electrons in $F_{16}$CoPc can be considered as fully localized and their contribution to transport ignored. From the measured Hall resistance and longitudinal resistivity we extract the values of the interfacial hole density ($n = 1.6 \cdot 10^{13}$ cm$^{-2}$) and mobility ($\mu = 1.2$ cm$^2$V$^{-1}$s$^{-1}$) for our $F_{16}$CoPc/rubrene interfaces.

With the exception of TTF-TCNQ,[5] the interfacial hole density in $F_{16}$CoPc/rubrene is the highest among all studied organic charge transfer interfaces.[3-13] The fact that the obtained value is higher than the one of F$_4$-TCNQ/rubrene interface ($1.0 \cdot 10^{13}$ cm$^{-2}$),[13] while the mobility values are very similar (F$_x$-TCNQ/rubrene interfaces: $\mu \sim 1.5$ cm$^2$V$^{-1}$s$^{-1}$),[13] is consistent with the lower resistivity of $F_{16}$CoPc/rubrene. Previous studies have shown that all F$_x$-TCNQ/rubrene interfaces have thermally activated behavior of the charge transport where the amount of charge transfer (carrier density) systematically increases with decreasing of the activation energy.[13] For F$_4$-TCNQ/rubrene the activation energy was found to be rather small ($E_a = 15$ meV, comparable to $k_B T$ at high measurement temperatures), which led to deviations from the thermally activated behavior in the high temperature range (i.e., between 200 and 300 K).[13] Following the trend found in the F$_x$-TCNQ/rubrene family and given the larger interfacial charge transfer in $F_{16}$CoPc/rubrene, quite pronounced deviations from a thermally activated behavior could be expected for the latter interface. Indeed, the results of temperature dependent transport measurements (see Figure 2b) show a decrease in resistivity upon cooling, indicating that transport at $F_{16}$CoPc/rubrene interfaces exhibit clear signatures of the intrinsic band-like regime, down to $T \sim 130$ K.

The mobility of holes propagating at the surface of rubrene crystals at $F_{16}$CoPc/rubrene interfaces ($\mu = 1.2$ cm$^2$V$^{-1}$s$^{-1}$) is much lower than in rubrene single-crystal field effect transistors (FETs).[32-35] It is also significantly lower than the value measured in rubrene FETs with polymeric gate dielectric whose dielectric constant has a value comparable to that typical of organic semiconductors.[33-35] The data show that the hole mobility measured $F_{16}$CoPc/rubrene interface is however comparable to the hole mobility in single-crystalline F$_x$-TCNQ/rubrene interfaces ($\mu \sim 1.5$ cm$^2$V$^{-1}$s$^{-1}$).[13] This finding may seem surprising, as one could have expected that the superior structural quality of F$_x$-TCNQ single crystals should have a positive influence on the mobility of charge carriers propagating at the interface. We believe that the explanation for the similar mobility observed in these interfaces is that the transferred electrons are in all cases localized (i.e., experimentally there is no sign of electron conductance neither in F$_x$-TCNQ/rubrene nor in $F_{16}$CoPc/rubrene ) and generate comparable potential fluctuations which are the cause for mobility-limiting disorder experienced by the holes propagating at the surface of rubrene[13] (why in F$_2$-TCNQ/rubrene electrons are localized[13] given the very high mobility of electrons in F$_2$-TCNQ single crystal FETs[30] remains to be understood).

To gain a better microscopic understanding of the energetics of $F_{16}$CoPc/rubrene interfaces, we have performed scanning Kelvin probe microscopy (SKPM)[36] experiments and have measured the contact potential difference between rubrene and $F_{16}$CoPc, which corresponds to the difference $\Delta E_F$ between the chemical potentials in the two materials. The measurements were performed on samples consisting of a 70 nm $F_{16}CoPc$ film evaporated onto a $SiO_2$ substrate, onto which a rubrene crystal was subsequently laminated. Measuring the difference in contact potential by scanning across the $F_{16}CoPc$/rubrene interface is particularly effective, because it enables the contact potential to be measured directly independently of the work function of the tip. A representative SKPM image and a line-scan contact potential measurement are shown in **Figure** 3a and 3b, respectively. The difference in the chemical potentials of the $F_{16}CoPc$ film and the rubrene crystal, $\Delta E_F$, can be extracted directly from the data and is found to be approximately 290 meV (see Figure 3b). This value is larger than the one obtained for $F_4$-TCNQ/rubrene, $\Delta E_F \sim 250$ meV, the largest in the $F_x$-TCNQ/rubrene family of interfaces.[13] Since a larger value of $\Delta E_F$ is normally conducive to a larger charge transfer, the outcome of SKPM experiments support the conclusions obtained from the transport measurements, namely that the charge transfer at the $F_{16}CoPc$/rubrene interface is larger than the charge transfer at any of interface of the $F_x$-TCNQ/rubrene family.

Finally, the electronic states of $F_{16}CoPc$/rubrene interfaces have been probed by photoemission spectroscopy in the valence as well as the core level region. Since straightforward photoemission spectroscopy on bulk rubrene crystals are prevented by charging effects,[37] measurements were performed on $F_{16}CoPc$/rubrene thin film interfaces. These interfaces were prepared using a gold (100) single crystal as a substrate, onto which a 5 nm rubrene film was deposited, followed by an $F_{16}CoPc$ film (more details on sample preparation and PES measurements can be found in the supporting information). Samples with different nominal thickness of the $F_{16}CoPc$ film ranging from 0.1 nm to 3.5 nm were investigated in order to identify particular changes that represent the interface region. From the continuous decrease of substrate core level intensities (see Figure S1 in the supporting information) we conclude that $F_{16}CoPc$ predominantly grows as subsequent layers onto the rubrene film surface. Moreover, the work function for the thickest $F_{16}CoPc$ film suggests that the $F_{16}CoPc$ molecules in thick films are essentially arranged in a standing geometry.[38]

**Figure** 4 summarizes the results of the photoemission studies of $F_{16}CoPc$/rubrene interfaces with different thicknesses of the $F_{16}CoPc$ layer. Panels a-c of Figure 4 depict the Co $3p_{3/2}$ core level emission spectra for three selected layer thicknesses. The spectrum obtained for the thick $F_{16}CoPc$ layer of 3.5 nm (Figure 4a) consists of a single, slightly asymmetric line which represents the two valent Co(II) in the center of $F_{16}CoPc$; the width and the shape of the spectral feature is determined by the Co $2p_{3/2}$ multiplet.[39] This result agrees perfectly with the spectra previously obtained for pure Co-phthalocyanine and Co-porphyrine molecules.[40-43]

In the case of thinner $F_{16}CoPc$ layers we observe changes in the Co $2p_{3/2}$ spectrum (see Figure 4b). Here, a second spectral feature appears at lower binding energies, which indicates a change in the valence of the Cobalt ion to Co(I) and corresponds to the $F_{16}CoPc$ molecules at the interface to rubrene. With further reducing the thickness of the $F_{16}CoPc$ film –i.e. increasing the contribution of the interfacial $F_{16}CoPc$ molecules to the measured signal– we see a clear increase of the relative intensity of the second feature (see Figure 4c). These observations are in good agreement with a number of studies where the interaction of Co-phthalocyanines with metal substrates has been reported.[40-42] The second feature in the spectrum arises due to a strong interaction at the corresponding interface that leads to a charge transfer and consequent reduction of the Co center in $F_{16}CoPc$.[40,41,44] Thus, our results demonstrate that at the $F_{16}CoPc$/rubrene interface the Co center of the $F_{16}CoPc$ is reduced due to a charge transfer from the rubrene molecules.

Photoemission spectra of the valence region of $F_{16}CoPc$ for three different layer thicknesses are presented in panels d-f of Figure 4. The data are fully consistent with the results obtained from the core level presented above. For a thick $F_{16}CoPc$ layer (Figure 4d) the spectrum consists of an emission line at about 1.2 eV binding energy arising from the highest occupied molecular orbital (HOMO) of $F_{16}CoPc$.[41] However for thinner layers, where the relative contribution to the signal from interfacial molecules is higher (Figure 4e, 4f), an additional feature appears at lower binding energy (about 0.75 eV). This feature can be associated to the $3d_{z^2}$ orbital of the phthalocyanines Co center[45] that gets filled due to the charge transfer from rubrene molecules and becomes therefore visible in PES. Consequently, the photoemission spectroscopy investigations complement our transport studies as they clearly indicate a charge transfer at the $F_{16}CoPc$/rubrene interface concomitant with a hole doping of rubrene. Moreover, they show that the Co ions in the neutral (before the transfer) and charged (after the transfer) $F_{16}CoPc$ molecules that coexist at the interface, have different electronic configurations and therefore different magnetic properties. Specifically, in the neutral $F_{16}CoPc$ molecules, Co(II) ion has $3d^7$ electronic configuration and a spin $S = ½$; after the charge transfer, the reduced Co(I) ion has the configuration $3d^8$ and $S = 0$.

The information drawn from the transport and spectroscopic measurements outline an interesting situation, namely the fact that $F_{16}CoPc$/rubrene interfaces support a high, band-like conductivity, with the charge carriers originating from interfacial charge transfer involving magnetic ions. These ingredients have not been established in any other organic charge transfer interface studied in the past. Having in mind the goal to induce a magnetic state in which interactions between molecular spins are mediated by charge carriers propagating from one magnetic molecule to the other, these ingredients are all necessary. In the case of $F_{16}CoPc$/rubrene interfaces, however, there is a serious limitation due to the fact that the electrons on the $F_{16}CoPc$ appear to be virtually completely localized and the transport is due to holes propagating on (non-magnetic) rubrene molecules. In principles, these holes can also mediate interaction between the magnetic $F_{16}CoPc$ molecules, as they are electronically coupled to them (that is why there is charge transfer). In practice, however, it seems obvious that the coupling is extremely small, making any cooperative magnetic phenomenon –if present at all– impossible to observe at realistic temperatures. These considerations clearly set the improvement of the coupling between charge carriers and magnetic moments of the molecules as a goal for future research, either on the molecular system considered here or on some analogous one. In this regard we emphasize that –even though in $F_{16}CoPc$/rubrene and in all $F_x$-TCNQ/rubrene interfaces electrons are localized– this is not an in-principle fundamental obstacle. Indeed, in other molecular systems electrons have been shown to delocalize: in PDIF-$CN_2$/rubrene interfaces, for example, conduction is fully mediated by electrons (no contribution of holes in rubrene to the conduction was detected)[8] and experiments in TTF/TCNQ interfaces are interpreted under the assumption that both electrons and holes are delocalized.[5]

In conclusion, we have performed systematic investigations of $F_{16}$CoPc/rubrene charge transfer interfaces by means of charge transport measurements, Hall effect, SKPM, XPS and UPS. We find that the charge transfer leads to significantly enhanced electrical conductivity and band-like transport, and we have determined the density, mobility and nature of charge carriers in the system (holes in rubrene). We also find that the amount of charge transfer in $F_{16}$CoPc/rubrene is high enough to cause the band-like transport in rubrene crystals at the interface. Finally, XPS and UPS measurements allow us to conclude that the charge transfer in $F_{16}$CoPc/rubrene fully involves the metal Co core of the phthalocyanine molecules making this system to be the first conducting organic interface in which charge transfer involves magnetic ions causing a change in their spin.

**Acknowledgements**
Y. K. and A. M. would like to thank A. Ferreira for technical support. Y. K. gratefully acknowledges the financial support from the German Research Foundation (DFG) through the Research Fellowships KR 4364/1-1 and KR 4364/2-1. A. M. gratefully acknowledges SNF fror financial support. F. R. and M. K. would like to thank R. Hübel, S. Leger and M. Naumann for technical assistance. Financial support by the German Research Foundation (DFG) within the Forschergruppe FOR 1154 (KN 393/14) is gratefully acknowledged.

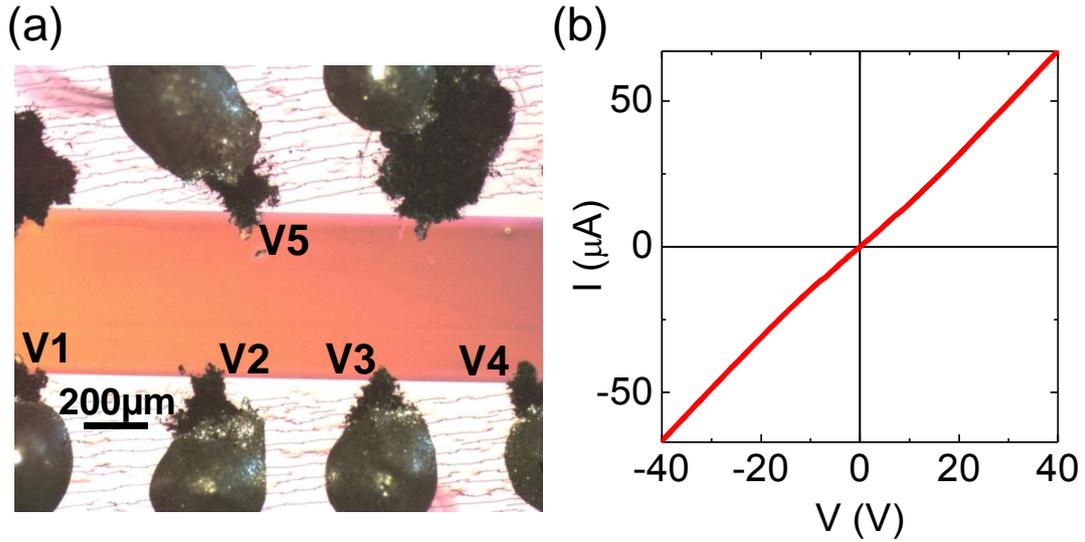

**Figure 1.** (a) Optical microscope image of a $F_{16}CoPc$/rubrene device. The rubrene single crystal is covered by a 70 nm $F_{16}CoPc$ film and contacted with conducting carbon paste. Contacts V1/V4 were used to source and drain current, contacts V2/V3 to measure the voltage and perform four-terminal resistance measurements, and conatcts V2/V5 for Hall voltage measurements. (b) Room temperature I-V curve for the $F_{16}CoPc$/rubrene device measured in a four-terminal configuration.

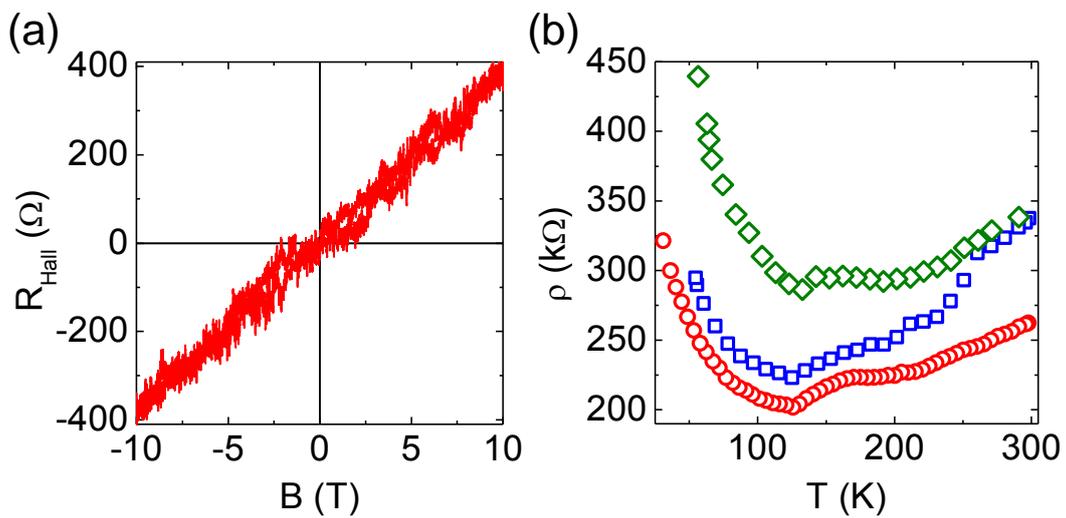

**Figure 2.** (a) Hall resistance vs. applied magnetic field measured at room temperature. (b) Temperature dependence of the resistivity of three different, nominally identical, $F_{16}CoPc$/rubrene devices.

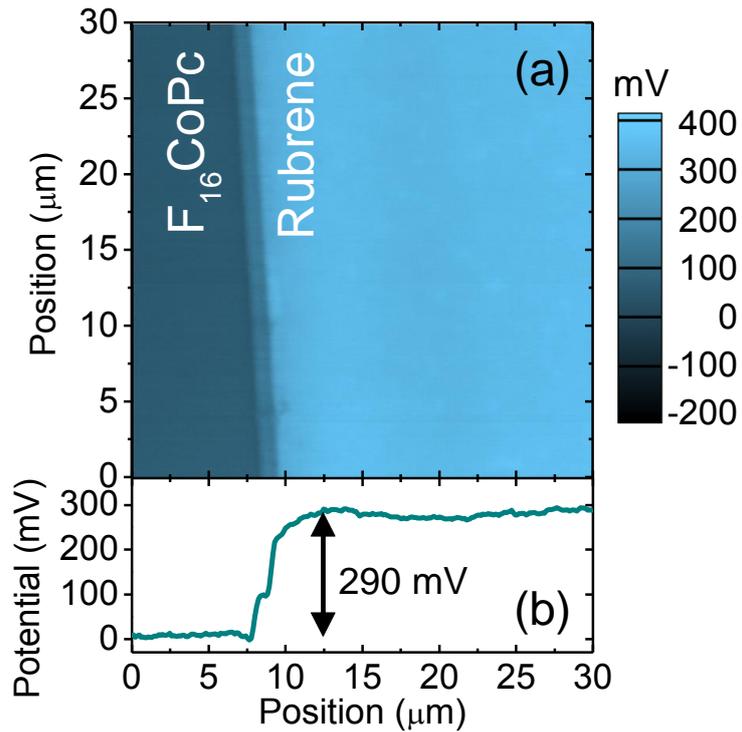

**Figure 3.** Scanning Kelvin probe microscopy measurements on an $F_{16}$CoPc/rubrene heterostructure. The results exhibit a clear step in both the topography (not shown) and in the contact potential (a) as the tip is moved from the surface of $F_{16}$CoPc film (left side of the images) to rubrene crystal (right side of the images). (b) Line-cut extracted from the SKPM image (a); the step corresponds to the difference in contact potential measured on the $F_{16}$CoPc and the rubrene.

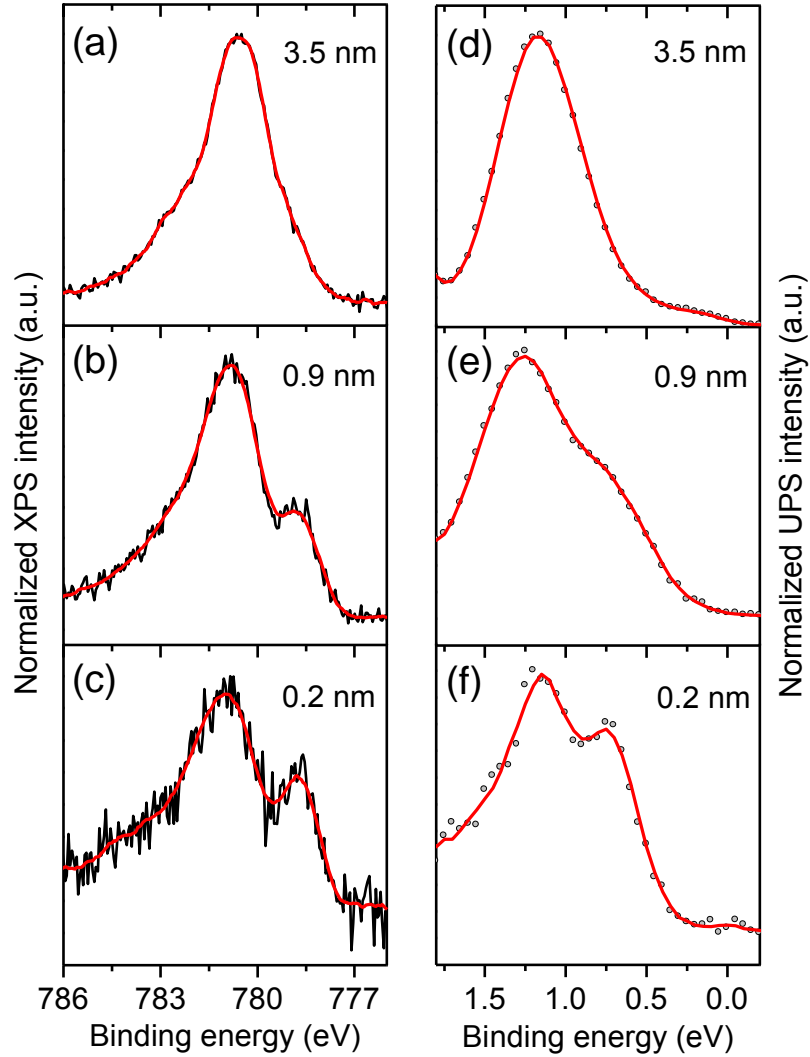

**Figure 4.** Left: Photoemission core level (XPS) spectra at the Co $2p_{3/2}$ core level of $F_{16}CoPc$/rubrene film hererostructure with different $F_{16}CoPc$ film thickness: 3.5 nm (a), 0.9 nm (b) and 0.2 nm (c). Additional feature related to the interfacial states appears in the spectrum of the thinner $F_{16}CoPc$ film. Right: Valence band photoemission (UPS) spectra of the valence region of $F_{16}CoPc$ with different film thickness: 3.5 nm (d), 0.9 nm (e) and 0.2 nm (f). The contribution of a pure rubrene film was subtracted from the spectra. The second peak in the spectrum of the thinner $F_{16}CoPc$ film corresponds to the $3d_{z^2}$ orbital of the phthalocyanines Co center that is empty in the normal state and gets filled due to the charge transfer from rubrene molecules.

# Supporting Information

**Charge transfer, band-like transport, and magnetic ions at $F_{16}CoPc$/rubrene interfaces**

*Yulia Krupskaya\*, Florian Rückerl, Martin Knupfer, and Alberto F. Morpurgo\**

1. **Photoelectron spectroscopy (PES) measurements**

The X-ray (XPS) and ultra-violet (UPS) photoelectron spectroscopy experiments were carried out using a two-chamber ultra-high vacuum system. The measurement chamber with a base pressure of about $2 \cdot 10^{-10}$ mbar is equipped with two light sources and an electron-energy analyzer PHOIBOS-150 purchased from SPECS. Besides a monochromatized Al $K_\alpha$ source with a Photon energy of 1486.6 eV to reach the deep lying core levels (XPS) the chamber also contains a He discharge lamp providing a Photon energy of 21.21 eV used to perform valence band measurements (UPS). The UPS experiments were carried out by applying a sample bias of -5 eV to obtain the exact secondary electron cutoff. Additionally, the recorded UPS spectra were corrected for the contribution of He satellite radiation. The total energy resolution of the spectrometer can be numbered to 0.35 eV for XPS and 0.15 eV for the UPS measurements.

2. **PES sample preparation**

A Gold (100) single crystal surface was used as substrate, which was prepared by repeated $Ar^+$ sputtering processes and annealing cycles. After this procedure a 5x20 surface reconstruction was observed by using low energy electron diffraction,[S1,S2] while no remaining contamination of the surface could be detected by monitoring with core level photoemission. Afterwards, a film of rubrene with a thickness of about 5 nm was prepared by an *in situ* thermal evaporation in the preparation chamber. The Au (100) single crystal substrate was kept at room temperature and the deposition rate was around 0.5 Å/min. To estimate and control the layer thickness during the deposition process the attenuation of the intensity of the prominent Au $4f_{7/2}$ substrate core level peak due to the deposition of the organic films was monitored, i.e. we used the procedure established by Seah and Dench.[S3]
After the rubrene layer was formed the $F_{16}CoPc$ molecules were deposited stepwise to produce various top-layer thicknesses starting from a thin (0.1 nm) and ending up with a thick (3.5 nm) layer of $F_{16}CoPc$ on top of the rubrene film.

3. **XPS at the Au 4f core level**

Figure S1 shows the spin orbit split Au 4f core level emission from the gold substrate for a pure rubrene film on top and for various thicknesses of $F_{16}CoPc$/rubrene film heterostructures. As expected for a subsequent layer growth of $F_{16}CoPc$ onto a rubrene film surface, the Au 4f core level spectra reveal a continuous attenuation of its intensity during the increase of the $F_{16}CoPc$ layer on top.

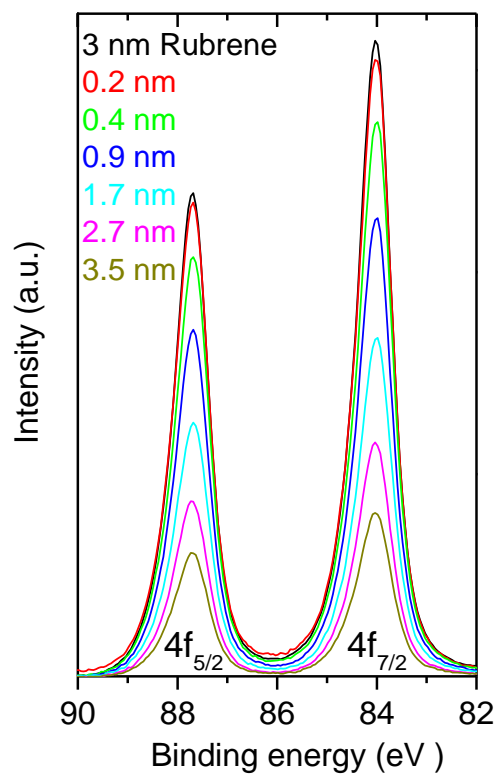

**Figure S1.** XPS spectra at the spin orbit split Au 4f core level of the gold substrate. The spectra represent a pure rubrene layer and different film thicknesses of the $F_{16}CoPc$/rubrene heterostructures. They exhibit a continuous, exponential-like decrease in intensity during the deposition of $F_{16}CoPc$. The data were taken with equal acquisition times.